\documentclass[fleqn,twoside]{article}
\usepackage{gc,amssymb}

\def\mn{_{\mu\nu}}

\def\mN{_\mu^\nu}

\def\cK{{\cal K}}

\def\Z{{\mathbb Z}}

\def\GR{general relativity}

\def\GR{general relativity}

\def\bw{brane world}

\def\oV{\overline{V}}
\def\vp{\overrightarrow{p}}

\heads  {S.T. Abdyrakhmanov, K.A. Bronnikov and B.E. Meierovich}
        {Uniqueness of RS2 type thick branes supported by a scalar field}

\begin{document}
\twocolumn[
\prepno{\bf gr-qc/0503055}{}
%%\Arthead{11}{2005}{1-2 (41-42)}{1}{5}

\bigskip
\bigskip
\bigskip

\Title {Uniqueness of RS2 type thick branes\yy supported by a scalar field}

\Aunames{Sergei T. Abdyrakhmanov\auth{a},
         Kirill A. Bronnikov\auth{a,b,1}
	 and Boris E. Meierovich\auth{c,2}}

\Addresses{
\addr a {Institute of Gravitation and Cosmology, PFUR,
        	6 Miklukho-Maklaya St., Moscow 117198, Russia}
\addr b {Centre for Gravitation and Fundam.
         Metrology, VNIIMS, 3-1 M. Ulyanovoy St., Moscow 117313, Russia}
\addr c {Kapitza Institute for Physical
                Problems, 2 Kosygina St., Moscow 117334, Russia}
}

\Abstract
   {We study thick \bw\ models as $\Z_2$-symmetric domain walls supported by
    a scalar field with an arbitrary potential $V(\phi)$ in 5D general
    relativity. Under the global regularity requirement, such configurations
    (i) have always an AdS asymptotic far from the brane, (ii) are only
    possible if $V(\phi)$ has an alternating sign and (iii) $V(\phi)$
    should satisfy a certain fine-tuning type equality. Thus a thick brane
    with any admissible $V(\phi)$ is a regularized version of the RS2 brane
    immersed in the AdS${}_5$ bulk. The thin brane limit is realized in a
    universal manner by including an arbitrary thick brane model in a
    one-parameter family, where the parameter $a$ is associated with brane
    thickness; the asymptotic value of $V(\phi)$ (related to $\Lambda_5$,
    the effective cosmological constant) remains $a$-independent. The
    problem of ordinary matter confinement on the brane is discussed for a
    test scalar field.  Its stress-energy tensor is found to diverge at the
    AdS horizon for both thin and thick branes, making a serious problem for
    this class of \bw\ models.	        }

]  %%%%%%%%%%%%%%%%%%%%%%%%
\email 1 {kb20@yandex.ru}
\email 2 {meierovich@yahoo.com}

\section{Introduction}

   According to the widely discussed \bw\ concept, the standard-model
   particles are confined on a hypersurface, called a brane, which is
   embedded in a higher-dimensional space called the bulk. This interesting
   and intriguing possibility from the viewpoint of the physical world
   outlook traces back to the 80s \cite{ARVSG}, and the present-day interest
   in it is largely related to the developments in superstring/M-theories
   \cite{Horava-Witten}. Various aspects of brane-world particle physics,
   gravity and cosmology are discussed in the recent review articles
   (\cite{Reviews}, see also references therein).

   Most of the studies are restricted to infinitely thin branes with
   delta-like localization of matter. This kind of models can, however, be
   only treated as an approximation since any fundamental underlying theory,
   be it quantum gravity or string theory, must contain a fundamental length
   beyond which a classical space-time description is impossible. It is
   therefore necessary to justify the infinitely thin brane approximation as
   a well-defined limit of a smooth structure, a thick brane, obtainable as
   a solution to coupled gravitational and matter field equations.

   In Ref.\,\cite{thick2} we, like many others, tried to describe a thick
   brane in the framework of 5D \GR\ as a domain wall separating two
   different states of a scalar field. Scalar field structures with
   arbitrary potentials were studied analytically, assuming $\Z_2$ symmetry
   with respect to the middle plane of the wall. The consideration was
   restricted to Poincar\'e branes, i.e., flat domain walls. A reason was
   that most of the existing problems are clearly seen even in these
   comparatively simple systems; moreover, in the majority of physical
   situations, the inner curvature of the brane itself is much smaller than
   the curvature related to brane formation, therefore the main qualitative
   features of Poincar\'e branes should survive in curved branes.

   This work confirmed and generalized the well-known results obtained in a
   number of specific models (see, among others, [6--10]):
%%   \cite{b1,soda,melfo,ghoroku,barcelo}:
   in the framework of 5D \GR, a globally regular thick brane always has an
   anti-de Sitter (AdS) asymptotic and is only possible if $V(\phi)$ has an
   alternating sign. In other words, it can be stated that {\it any\/}
   regular thick Poincar\'e brane supported by a scalar field is a smooth
   version of the well-known Randall-Sundrum second model (RS2) \cite{RS2}.

   The problem of a well-defined thin brane limit was also discussed in
   \cite{thick2} and was solved partly, using as an example exactly
   solvable models with scalar field potentials consisting of two constant
   steps. The RS2 limit, with the corresponding (fine-tuned) relation
   between the brane tension $\sigma$ and the bulk cosmological constant
   $\Lambda_5$, was found to be independent of the two shape factors of the
   potential (the steps' relative height and thickness) which were kept
   constant in the transition to zero thickness. Examples of similar
   transitions for other specific potentials have also been found previously
   \cite{melfo,ghoroku}

   In this paper we re-consider this problem and indicate a simple algorithm
   which, given any regular thick brane model, includes it into a
   one-parameter family of similar models with a parameter $a$ related to
   brane thickness. The limit $a\to 0$ corresponds to the RS2 model. We thus
   prove the existence of a correct thin brane limit in a universal way for
   thick brane models with potentials of arbitrary shape.

   In the concluding section 5 we also briefly discuss the problem of
   ordinary matter confinement using as an example a test scalar field.
   It turns out that its stress-energy tensor diverges on the AdS horizon,
   this feature being common for a thin RS2 brane and all its thick
   counterparts.

\section{Field equations and boundary conditions}

   We consider \GR\ in a 5D space-time, where we distinguish the usual
   four coordinates $x^\mu$, $\mu = 0,1,2,3$ and the fifth coordinate
   $x^4 = z$, to be used for describing the direction across the brane.
   The action is taken in the form $S = \int \sqrt{g} L\,d^5 x $, where
   $g = |\det (g_{AB})|$, $A,B = 0,1,2,3,4$ and $g_{AB}$ is the 5D
   metric tensor; the Lagrangian density has the form
\beq
    L = \frac{R}{2\kappa^2} + L_\phi, \cm
               L_\phi = \Half \d^A \phi\, \d_A\phi - V(\phi),    \label{L}
\eeq
   ($R$ is the 5D Ricci scalar and $\kappa$ the 5D gravitational constant)
   and leads to the Einstein-scalar equations
\bearr                                                          \label{EE}
   R_{A}^{B} - \half\delta_{A}^{B} R = -\kappa^2 T_{A}^{B}
                    = -\kappa^2 [\d_A\phi \d^B\phi - \delta_A^B\, L_\phi],
\yyy                                                          \label{eq_phi}
     \d_A (\sqrt{g}\, g^{AB}\d_B \phi) = - \sqrt{g}\, d V/d\phi
\ear
   where \eq (\ref{eq_phi}) is a consequence of (\ref {EE}) due to the
   Bianchi identities.

   We seek regular solutions describing a static domain wall (thick
   Minkowski brane), possessing $\Z_2$ symmetry with respect to the
   hypersurface $z=0$.

    Consider \eqs (\ref{EE}), (\ref{eq_phi}) for $\phi = \phi (z)$
    and the metric
\beq                                                        \label{ds_gen}
     ds_5^2 = \e^{2F(z)} \eta\mn dx^\mu dx^\nu - \e^{8F(z)} dz^2,
\eeq
    where $\eta\mn$ is the 4D Minkowski metric and we have chosen the
    harmonic coordinate $z$, such that $\sqrt{g} g^{zz}=-1$.
    As a result, the 5D Ricci tensor acquires an especially simple form:
\bear                                                    \label{R_AB-harm}
     R_0^0 \eql R_1^1 = R_2^2 = R_3^3 = -\e^{-8F} F'',
\nn
     R_z^z \eql -4 \e^{-8F} (F'' - 3{F'}^2)
\ear
    The Kretschmann scalar ${\cal K} = R_{ABCD} R^{ABCD}$ is
\beq
      {\cal K} = 4\bigl[ \e^{-5F}(\e^{-3F} F')'\bigr]^2         \label{Kr}
                + 6\bigl( \e^{-4F} F'\bigr)^4.
\eeq
    For the metric (\ref{ds_gen}), $\cK$ is a sum of squares of all nonzero
    components $R_{AB}{}^{CD}$ of the Riemann tensor, therefore its finite
    value is necessary and sufficient for finiteness of all algebraic
    curvature invariants.

    The 5D Einstein equations (\ref{EE}) in our case reduce to
\bearr                                                      \label{F''=}
      F'' = -\frac{2\kappa^2 }{3}\e^{8F}V,
\yyy
      3(-F'' + 4 {F'}^2) = \kappa^2{\phi'}^2,               \label{no_V}
\yyy                                                        \label{1int}
      {F'}^2 = \frac{\kappa^2 }{6}\left( \Half{\phi'}^2 -\e^{8F}V\right),
\ear
    where (\ref{1int}) is a first integral of the other two equations.
    The scalar field equation (\ref{eq_phi}) reads
\beq                                                    \label{phi''=}
            \phi'' = \e^{8F} d V/d\phi
\eeq
    and is also a consequence of (\ref{F''=}) and (\ref{no_V}). \eqs
    (\ref{F''=}) and (\ref{1int}) may be taken as a complete set of
    equations for $F(z)$ and $\phi (z)$; it is third-order and requires
    three boundary conditions.

    Now, the $\Z_2$ symmetry assumption dictates the boundary condition
    $F'(0) =0$. Then, assigning $F(0) =0$ by a proper choice of
    the time scale, we arrive at an unambiguous value of ${\phi'}^2$:
    ${\phi'}^2 (0) = 2V(0)$. So $F(z)$ is an even function while $\phi (z)$
    is an odd one. A complete set of boundary conditions
    at $z=0$, compatible with $\Z_2$ symmetry, is
\beq                                                      \label{Bound2}
    F(0) = F'(0) = 0,  \qquad  \phi (0) = 0.
\eeq
    Thus for any fixed function $V(\phi)$, there is no free parameter in the
    equations and boundary conditions, i.e., the solution is already
    uniquely determined. The requirement of regularity at large $z$ should
    thus lead to an additional constraint on the function $V(\phi)$, leading
    to a ``fine tuning'' between the brane and bulk parameters
    \cite{Shiromizu00}.

\section {Regularity conditions}

    Let us briefly reproduce the reasoning of Ref.\,\cite{thick2} on
    regularity of the solutions at the asymptotic $z \to \infty$.

    Regularity of the metric [see (\ref{Kr})] implies $|b'(z)| < \infty$
    where $b(z) \equiv \e^{-4F}$. Regularity of the geometry leads, via the
    Einstein equations, to finiteness of certain scalar field
    characteristics. On the whole, we should require
\beq                                                            \label{reg}
    |b'(z)| <\infty, \quad\  |V(\phi)| < \infty, \quad\
        				b(z)|\phi'(z)| < \infty.   %% 12
\eeq

    \eq (\ref{no_V}) leads to $b''(z) > 0$. Since $b'(0) = 0$, this means
    that $b(u)$ increases at $z > 0$ and inevitably grows to infinity at
    large $z$ at least linearly for any nontrivial solution.
    The growth is precisely asymptotically linear due to (\ref{reg}), $b'
    \to \const >0$, and hence $F' \approx  -1/(4z)$ at large $z$.

    Returning to (\ref{F''=}) and integrating, we obtain
\beq                                                          \label{oV_}
    F'(z) = \oV(z) := -\frac{2}{3}\kappa^2
                              \int_{0}^{z}\e^{8F} V\,dz.           %% 13
\eeq
    For regular solutions we necessarily have
\bearr                                                   \label{oV}
    \oV (\infty) =0.                                               %% 14
\ear
    This is the above-mentioned fine-tuning condition in terms of the
    potential $V$. The integral $\oV(z)$ is the invariant full potential
    energy per unit 3-volume in the layer from zero to $z$. Since
    $\e^{8F} = 1/b^2 > 0$, a nontrivial potential $V(\phi)$ must change its
    sign at least once to yield $\oV (\infty) = 0$.

    It is easy to show that in regular solutions $\phi' = o(1/z)$.
    Therefore, $\e^{-4F}\phi' \to 0$ at large $z$, and \eq
    (\ref{1int}) shows that $V$ tends to a finite negative value. If
\beq                                                           \label{F-as}
     b(z)=\e^{-4F} \approx kz, \quad\ k = \const > 0, \quad\ z\to \infty,
\eeq
    then
\beq
    \kappa^2 V\Big|_{z=\infty} = \Lambda_5 =- \frac{3k^2}{8},  \label{V-as}
\eeq
    where $\Lambda_5$ is the effective cosmological constant.

    Thus $V(\phi)$ changes its sign at least once, tends to a negative
    value as $z\to \infty$, and the integral $\oV(\infty)$ is zero.
    If $\phi$ tends to a finite value $\phi_\infty$, $V(\phi)$ has a minimum
    at $\phi=\phi_\infty$.

    This behaviour is not unique: despite $\phi' = o(1/z)$, one cannot
    exclude that $\phi \to \infty$, though slower than $\ln z$ (it can be,
    for instance, $\phi' \sim 1/(z\ln z)$ and $\phi \sim \ln\ln z)$. Then
    $V (\phi)$ tends to the value (\ref{V-as}) as $\phi \to \infty$.

    In any case, due to oddness of $\phi (z)$, the values $\phi (+\infty) =
    -\phi(-\infty) \ne 0$ make the domain wall topologically stable. The
    postulated $\Z_2$ symmetry implies that $V(\phi)$ is an even function.
    Its asymptotic value, $V(z = \pm\infty) = \Lambda_5/\kappa^2 < 0$ plays
    the role of a cosmological constant at the bulk asymptotic, and the
    metric is asymptotically anti-de Sitter (AdS). The values $z = \pm\infty$
    then correspond to an anti-de Sitter horizon.

\section{Thin brane limit}

    Consider the thin brane limit of regular solutions with finite
    $\phi_\infty$, leaving aside the above case of slowly growing $\phi(z)$.
    We thus have $V(\phi)$ with a minimum at $\phi = \phi_\infty$, and there
    holds the ``fine tuning'' condition $\oV(\infty) =0$.

    The action $S$ with the Lagrangian $L = L_G + L_\phi$
    may be split into the bulk and brane parts,
\bear  \nq                                                     \label{S_tot}
    S \eql S_{\rm bulk} + S_{\rm brane},
\\     \nq                                                    \label{S_bulk}
    S_{\rm bulk} \eql -\int \frac{R - 2\Lambda_5}{2\kappa^2 }
     \sqrt{g}d^{5}x,\quad\  \frac{\Lambda_5}{\kappa^2} = V(\phi_\infty),
\\     \nq                                                   \label{S_brane}
    S_{\rm brane} \eql \int \!\left[ \frac{1}{2}
           \d_A\phi\, \d^A\phi + V(\phi_\infty) -V(\phi) \right]
               \! \sqrt{g}\,d^{5}x.
\ear
    The brane action may be presented in the form
\bearr                                                      \label{tension}
     S_{\rm brane} = - \int \sigma \,d^4 x,
\nnn
    \sigma = \int_{-\infty}^\infty \left[- \frac{1}{2}\d_A\phi\, \d^A\phi
                    + V(\phi) - V(\phi_\infty) \right] \e^{8F} dz,
\ear
    where the quantity $\sigma$ can be regarded as the brane tension. It is
    equal to the total scalar field energy per unit 3-volume on the brane,
    in which the potential energy is counted from the vacuum level
    $V_\infty$.

    Randall and Sundrum's second model (RS2) of a thin brane \cite{RS2} is
    based on the splitting (\ref{S_tot}): assuming a delta-like matter
    distribution characterized by the tension $\sigma$ and using the
    Israel matching condition for the 5D metric, they found the
    fine-tuning condition
\beq                                                      \label{RS_fine}
     6 \Lambda_5 =  -\kappa^4\sigma^2.
\eeq

    In our approach, a transition to a thin brane can be carried out along a
    family of solutions with different potentials but a fixed value of
    $\Lambda_5$, which determines a length scale $1/\sqrt{\Lambda_5}$ in the
    bulk, independent of the brane thickness. This corresponds to a brane as
    a domain wall between two vacua with equal and fixed energy densities.

    Then, if the thin brane concept is correct, we should expect that,
    independently of the specific form of the potential,
\beq                                                      \label{thin_lim}
    \lim\limits_{a \to 0} \frac{|\Lambda_5|}{\kappa^4 \sigma^2} = \frac 16,
\eeq
    the parameter $a$ characterizing the brane thickness.

    To consider the problem, let us introduce more convenient variables
\beq                                                       \label{def_fv}
       f(z) := \frac{2\kappa}{\sqrt{3}} \phi(z), \cm
       v(f) := \frac{8}{3} \kappa^2 V(\phi).
\eeq
    \eqs (\ref{F''=})---(\ref{1int}) are then rewritten as
\bear
     v \eql bb'' - b'^2 \,=\, b^2 f'^2 - b'^2,               \label{v(z)}
\\
        f'^2 \eql b''/b.                                    \label{f'(z)}
\ear
    The boundary conditions at $z = 0$ are
\beq
    b(0) = 1, \cm b'(0) = 0, \cm f(0) = 0.                \label{bound_0}
\eeq

    Let there be any function $b(z)$ describing a thick brane
    in an AdS background. This means that  $b(z)$ satisfies
    \eqs (\ref{v(z)})---(\ref{f'(z)}) and the following boundary conditions
    at $z=0$  (location of the brane) and $z\to \infty$ (AdS horizon):
\bearr
       b(0) = 1, \cm b'(0) = 0,
\nnn
       b(z) \approx kz + \const \cm (z\to\infty)           \label{bound_1}
\ear
    so that the potential $v \to -k^{2} =\const$ ($k >0$) as $z\to \infty$.
    The tension $\sigma$ is expressed as
\beq
       \sigma =  \frac{3}{8\kappa^2}                           \label{sig}
       		\int_{-\infty}^{\infty} dz\,
			\biggl(\frac{b''}{b} + \frac{k}{b^2}\biggr).
\eeq

    A well-defined thin brane limit means that this thick-brane solution
    should be included in a family of solutions with a parameter $a$ (to be
    interpreted as a thickness parameter), say, $B(a,z)$ such that
    $B(1,z) = b(z)$, which should satisfy the following requirements:

\medskip\noi
    1) At each fixed $a > 0$, the function $B(a,z)$ should satisfy \eqs
    (\ref{bound_1}) with the same constant $k$, i.e., it should be a
    thick-brane solution with the same cosmological constant
    $\Lambda_{5} = -\frac{3}{8}k^{2}$. (The potential $v$ should evidently
    be $a$-dependent since its limiting form should be $v = -k^2$ at all
    $z\ne 0$.)

\medskip\noi
    2) At each fixed $z \ne 0$, the function $B(a,z) \to 1+k|z|$ as $a\to 0$,
    i.e., the metric should tend to the AdS metric with the properly chosen
    time scale at $z = 0$.

\medskip
    Then the limiting solution corresponds to the RS2 thin brane. In
    particular, the relation between the brane tension $\sigma$ and
    $\Lambda_5$ should be as in the RS2 model, i.e., (\ref{RS_fine}), or
\beq
    	\sigma = 3k/(2\kappa^2).                            \label{sig-lim}
\eeq

    The following function $B(a,z)$ evidently satisfies the above
    requirements 1) and 2):
\beq
       B(a,z) = 1 - a + ab\left( \frac{z}{a}\right),              \label{B}
\eeq
    where $b()$ means the functional dependence specified in the original
    function $b(z)$. Substituting $B(a,z)$ instead of $b(z)$ into
    (\ref{sig}) and considering the limit $a\to 0$, we see that the second
    term of the integrand yields precisely half the desired value
    (\ref{sig-lim}), namely, $3k/(4\kappa^2)$. The other half is given
    by the first term, since, in the limit $a\to 0$, we have
    $B'' \to 2k \delta(z)$.

    It is of interest to note that only half of the RS2 brane tension
    is related to an energy concentrated on the brane and described by the
    first term in (\ref{sig}), while the other half is distributed in the
    bulk proportionally to $(1 + k|z|)^{-2}$.

    The limiting form of the potential is
\beq
	v(z)\Big|_{a=0} = 2k \delta(z) -k^2.                 \label{v-lim}
\eeq

\section {Concluding remarks. The confinement problem}

    We have continued a general study of regular domain walls (thick branes)
    supported by a minimally coupled scalar field with an arbitrary
    potential in 5D \GR. It has been previously shown \cite{thick2} that the
    {\it only\/} kind of asymptotic for such walls is AdS, that the
    potentials $V(\phi)$ able to create such configurations have an
    alternating sign and satisfy the fine tuning condition (\ref{oV}).
    We have now confirmed in a general form that the zero thickness limit
    of such branes is well defined and conforms to the RS2 \bw\ model.

    The family of solutions (\ref{B}), derived from any appropriate solution
    (which is entirely characterized by the function $b(z)$), is probably
    the simplest possible family that realizes the thin brane limit, but it
    is evidently not unique.

    Explicit examples of a transition to thin branes with some special
    potentials $V(\phi)$ were studied previously (see \cite{melfo,ghoroku}
    and references therein) with the same result.

    It should be stressed, however, that, to be considered as models of our
    Universe, \bw\ models like those discussed here must satisfy two major
    requirements: (i) ordinary matter should be confined to the brane to
    account for the fact that extra dimensions are not observed, and (ii)
    Newton's law of gravity should be reproduced on the brane in a
    non-relativistic limit. These issues, which have already been treated in
    a number of papers (among others, \cite{Reviews,b1,soda,b2})
%%{rub_dub,middleton,seahra},
    turn out to be quite nontrivial, and we hope to discuss them in future
    publications. Let us briefly illustrate the confinement problem, using
    as an example a test scalar field $\chi$ in the background of a
    thick brane with the metric (\ref{ds_gen}).

    Consider a scalar field $\chi$ with the Lagrangian
\beq                                                           \label{L_hi}
    L_{\chi } = \half \d_A \chi \d^A \chi - \half m_0^2\chi^* \chi
    		+ \half \lambda \phi^2 \chi^{\ast }\chi,
\eeq
    where $\chi^*$ is the complex congugate field, and the last term
    describes a possible interaction between $\chi$ and the wall scalar
    field $\phi$; $\lambda$ is a coupling constant. The influence of $\chi$
    on the wall structure is neglected. The field $\chi (x^{A})$, which may
    be interpreted as the wave function of a $\chi$-particle, satisfies the
    Klein-Gordon-Fock equation
\beq                                        	        	\label{KG}
    \frac{1}{\sqrt{g}}\d_{A}\left( \sqrt{g}g^{AB}
                    \d_{B}\chi \right) = (\lambda \phi^2 - m_0^2) \chi ,
\eeq
    which is linear and homogeneous with respect to $\chi$. Its coefficients
    depend on $z$ only, while the coordinates $x^{\mu}$ are cyclic
    variables. The canonically conjugate momenta $p_{\mu }= (E,\,\vp)$ are
    integrals of motion, and we can present $\chi (x^{A}) $ in the form
\beq                                                            \label{sepa}
    \chi (x^{A}) = X(z) \exp (-ip_{\mu }x^{\mu }),
    		   				\qquad \mu =0,1,2,3.
\eeq
    The function $X (z)$ determines the $\chi$ field distribution across
    the brane and satisfies the linear homogeneous equation
\beq                                                      	\label{hi''}
    X'' + \sqrt{g} (p^{\mu }p_{\mu } + \lambda \phi^2 -m_{0}^2 )
            X = 0.
\eeq

    We can consider a $\chi$-particle to be localized on the brane if
    the stress-energy tensor (SET) of the $\chi$ field, $T\mN [\chi]$, is
    finite in the whole 5-space and decays sufficiently rapidly at large
    $z$. If $T\mN [\chi]$ somewhere tends to infinity, this evidently
    violates the test field assumption.

    As an evident necessary condition of localization, one can require
    converging $\chi$ field energy per unit 3-volume of the brane, i.e.,
\bearr                                                       \label{E_tot}
     E_{\rm tot}[\chi] = \int_{-\infty}^{\infty} T^t_t \sqrt{g} dz
\nnn \quad
     = \int_{0}^{\infty}
            \e^{8F}\Bigl[ \e^{-2F}(E^2 + \vp^2)X^2
\nnn \cm
	    + (m_0^2 -\lambda\phi^2)X^2 + \e^{-8F} X'^2 \Bigr] dz < \infty.
\ear
    The inequality (\ref{E_tot}) evidently implies a finite norm of the
    $\chi$ field defined as
\beq	         					 \label{hi-norm}
    \|\chi\|^2 = \int_{-\infty }^{\infty }\sqrt{g}\,\chi^*\chi\,dz
    	       = \int_{-\infty }^{\infty } \e^{-8F}\,X^2 \,dz.
\eeq

    \eq (\ref{hi''}) is rewritten as
\beq                                                  \label{hi''curv}
    \chi'' + \left[ \e^{6F} (E^2 -\vp ^2)
                   +\e^{8F} (\lambda\phi^2 -m_0^2) \right] \chi = 0.
\eeq
    The term $\e^{8F}( \lambda\phi^2 -m_{0}^2) $ describes interaction of a
    $\chi$-particle with the brane. If $\lambda = 0$, it is purely
    gravitational, while $\lambda\ne 0$ describes an additional,
    non-gravitational interaction between $\phi$ and $\chi$.

    Recall now that the space-time regularity requirement leads to
    $\oV(\infty) =0$, and $\e^{4F} \sim 1/z$. The term $\sim \e^{8F}$ in \eq
    (\ref{hi''curv}) vanishes faster than the one with $\e^{6F}$, and the
    equation determining the behavior of $\chi$ at large $z$ is
    $X'' + \e^{6F}(E^2 -\vp ^2 ) X =0$, or, due to (\ref{F-as}),
\bearr
    	X'' + \frac{\vartheta }{z^{3/2}}X =0,            \label{hi''as}
\nnn \cm
    	\vartheta^2  = (E^2 {-} \vp^2)
    	   \left[ 2\sqrt{\fract 23 \kappa | V(\phi_\infty)|}\right]^{-3/2}.
\ear
    Its asymptotic solution is
\beq                                                 	\label{hi-as}
      X = Cz^{3/8} \sin (4\vartheta z^{1/4} + \varphi_0 ),
\cm
		    z\to \infty,
\eeq
    where $C$ and $\varphi_0 $ are integration constants. We see that the
    wave function (\ref{hi-as}) not only does not vanish as $z\to \infty $,
    but oscillates with an increasing amplitude. As a result, the SET
    components $T\mN [\chi]$ are infinite at $z=\infty$, i.e., at the AdS
    horizon. Moreover, the integral (\ref{E_tot}) diverges since, due to the
    term proportional to $X'^2 \sim z^{-3/4}$, it behaves as $\int
    z^{-3/4}\,dz$. Meanwhile, the normalization integral (\ref{hi-norm})
    converges since the integrand behaves as $z^{-5/4}$. The latter result
    is sometimes treated as a sufficient condition for localization, but, in
    our view, it is not the case since the very existence of the brane
    configuration is put to doubt if the test field SET is somewhere
    infinite.

    The origin of the above divergence is the gravitational field which
    repels matter from the brane and pulls it to the AdS horizon. The
    universal AdS asymptotic leads to the solution (\ref{hi-as}) whose form
    is insensitive to the test field mass and to its possible interaction
    with the brane-supporting field $\phi$.

    For this reason we think that this class of models is unsatisfactory
    from a physical viewpoint. One of the ways in a search for \bw\
    models free from this problem is to consider higher-dimensional bulks,
    and we hope to discuss them in our future publications.

\Acknow
    {KB acknowledges partial financial support from ISTC Proj. \#\,1655.}

\end{document}